\documentclass[conference]{IEEEtran}
\IEEEoverridecommandlockouts
\usepackage{cite}
\usepackage{amsmath,amssymb,amsfonts}
\usepackage{algorithmic}
\usepackage{graphicx}
\usepackage{textcomp}
\usepackage{xcolor}
\usepackage{caption}
\usepackage{booktabs}
\usepackage{url}
\usepackage{subfigure}
\usepackage{tcolorbox}
\usepackage{balance}
\usepackage{multirow}
\def\BibTeX{{\rm B\kern-.05em{\sc i\kern-.025em b}\kern-.08em
    T\kern-.1667em\lower.7ex\hbox{E}\kern-.125emX}}

\usepackage{datenumber}

\definecolor{chestnut}{rgb}{0.8, 0.36, 0.36}

\definecolor{chestnut}{rgb}{0.8, 0.36, 0.36}

\newcounter{dateone}\newcounter{datetwo}%
\newcommand{\daydifftoday}[3]{%
\setmydatenumber{dateone}{\the\year}{\the\month}{\the\day}%
\setmydatenumber{datetwo}{#1}{#2}{#3}%
\addtocounter{datetwo}{-\thedateone}%
\thedatetwo
}

\usepackage{graphicx} 
\graphicspath{{pic/}} 

\begin{document}
\title{Visualizing Contributor Code Competency for \\ PyPI Libraries: Preliminary Results
}
\author{\IEEEauthorblockN{ Indira Febriyanti}
\IEEEauthorblockA{\textit{Universitas Muhammadiyah Surakarta}\\
l200194197@student.ums.ac.id}
\and
\IEEEauthorblockN{Raula Gaikovina Kula}
\IEEEauthorblockA{\textit{Nara Institute of Science and Technology}\\
raula-k@is.naist.jp}
\and
\IEEEauthorblockN{Ruksit Rojpaisarnkit}
\IEEEauthorblockA{\textit{Nara Institute of Science and Technology}\\
rojpaisarnkit.ruksit.rn1@is.naist.jp}
\and
\IEEEauthorblockN{Kanchanok Kannee}
\IEEEauthorblockA{\textit{Nara Institute of Science and Technology}\\
kanchanok.kannee.kg8@is.naist.jp}
\and
\IEEEauthorblockN{Yusuf Sulistyo Nugroho}
\IEEEauthorblockA{\textit{Universitas Muhammadiyah Surakarta}\\
yusuf.nugroho@ums.ac.id}
\and
\IEEEauthorblockN{Kenichi Matsumoto}
\IEEEauthorblockA{\textit{Nara Institute of Science and Technology}\\
matumoto@is.naist.jp}
}

\maketitle

\begin{abstract}
Python is known to be used by beginners to professional programmers.
Python provides functionality to its community of users through PyPI libraries, which allows developers to reuse functionalities to an application.
However, it is unknown the extent to which these PyPI libraries require proficient code in their implementation.
We conjecture that PyPI contributors may decide to implement more advanced Pythonic code, or stick with more basic Python code.
Are complex codes only committed by few contributors, or only to specific files?
The new idea in this paper is to confirm who and where complex code is implemented.
Hence, we present a visualization to show the relationship between proficient code, contributors, and files.
Analyzing four PyPI projects, we are able to explore which files contain more elegant code, and which contributors committed to these files.
Our results show that most files contain more basic competency files, and that not every contributor contributes competent code. 
We show how~our visualization is able to summarize such information, and opens up different possibilities for understanding how to make elegant contributions.
\end{abstract}
\begin{IEEEkeywords}
Code Competency, Pythonic Code, Software Visualization
\end{IEEEkeywords}

\section{Introduction}

Python is known to be versatile, suitable for a wide range of people, from beginners who want to create small scripts to professional programmers.
It is also used by many developers not having a formal computer science education, for example, in the case of the BioPython collection~\cite{cock2009biopython}.
Much of science-related software is developed in Python, and much of it is by non-IT professionals.
Beyond a programming language, Python is a live community that has its \emph{culture}.
An important part of this culture is about how to conceive solutions for certain problems.
These ideas are portrayed in the ``The Zen of Python''\footnote{https://www.python.org/dev/peps/pep-0020/}. 
In this respect, it can be said that some solutions are considered more \emph{elegant} than others,
the most elegant being the most proficiently \emph{Pythonic}~\cite{alexandru2018usage}.
Other works have shown how there is indeed more than one way to code in Python~\cite{10.1145/3486608.3486909}.
However, it is unsure the extent to which developers actually use more elegant solutions (proficient) in their code. 

Much like other large programming language communities, the Python community has its own set of community contributed libraries, that any programmer can reuse in their applications. 
This is known as the Python Package Index (PyPI) libraries\footnote{\url{https://pypi.org/}} that is aimed to help Python developers. 
Due to the open source nature of these libraries, they also face becoming obsolete if they are unable to attract contributors to help with the maintenance of these libraries.
Although these libraries are highly relied upon by the community, we are unsure whether or not sure of the competency required for developers to maintain, especially in terms of the contributions, and the extent to which they are elegant.

Studies of proficiency through code analysis have existed for a number of works. The programmer uses elements of code based on their experiences with the language and knowledge of its elements. Some elements of code are easier to understand than others and some elements of code may indicate the skills and experiences of the programmer. Robles et al.~\cite{robles2022pycefr} built a detection tool to detect and measure the competency level of each code element of Python language in six levels of Python proficiency categories. They categorized these six levels depending on the proficiency required to create and understand the code element. Phan-Udom et al.~\cite{phanudom2020teddy} develop an automated tool for checking the python idiom usage and provide the recommendation of using python idiom during code review.
Other related works involve exploring Pythonic usage over time ~\cite{Sakulniwat}, and their performance~\cite{leelaprute2022does}.
Recently, there has been interesting work into how to make Python code more Pythonic through refactoring techniques~\cite{zhang2022making}.

\begin{table*}
    \centering
    \caption{Overview of the projects used in the study}
    \label{tab:overviewproject}
    \begin{tabular}{llrrrrr}
        \toprule
        Analyzed project & GitHub Webpage & \# dependencies & \# stars & \# contributors & \# files & \# commits\\
        \midrule
        Requests     & \url{https://github.com/psf/requests} & 41.2K & 48.2K & 612 & 31 & 6,126\\
        Flask        &\url{https://github.com/pallets/flask} & 5.5K & 60.4K & 678 & 62 & 4,839\\
        Jinja2       &\url{https://github.com/pallets/jinja} &  6.16K & 8.7K & 267 & 60 & 2,645\\
        Pytz        &\url{https://github.com/stub42/pytz} & 4.41K & 220 & 15 & 15 & 614 \\
        \bottomrule
    \end{tabular}%
    
\end{table*}

\begin{figure}[t]
    \centering
    \includegraphics[width=0.25\textwidth]{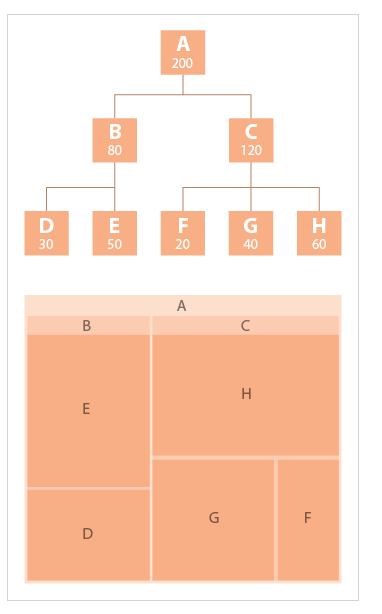}
    \caption{Basic Concept of the Treemap Charts transformed from a tree to area size.}
    \label{fig:hierarchy}
\end{figure}

To understand the extent by which more advanced Python code is implemented, in this early results paper we present a visualization showing the relationship between competent code and contributors.  
In this preliminary study, we visualize the competency of contributor commits in the GitHub repository. 
Our study covers four PyPI libraries, with a total of 149 files and 1,071 contributors. 
For the visualization,  we used a tree dependency (i.e., Treemap Charts). 
Based on the key goal of the research, we ask two research questions: 
\begin{itemize}
    \item \textit{RQ1) To what extent is proficient code prevalent in a project?}
    \item \textit{RQ2) To what extent which contributors commit competent code?}
\end{itemize}

From our visualization, we show the two results.
First, our visualization shows that Basic Competency (A \& B) is prevalent in PyPI projects, but also contains a much smaller percentage of proficient code.
Second, our visualization shows that most contributors still contribute basic competency code compared to proficient code. 
Our visualizations are available at \url{https://github.com/NAIST-SE/PyPICompViz}.




\section{Using a Visualization Tree dependency}

As shown in Figure~\ref{fig:hierarchy}, we adopt the Treemap charts, as it is an alternative to show dependencies between different entities. 
This way we can show the relationship between contributors, and the competency of codes in a file. 
Hence these relationships can be characterized as different categories.

Treemaps are an alternative way of visualizing the hierarchical structure of a Tree Diagram while also displaying quantities for each category via area size~\cite{9604836,9373115}.
In detail, each category is assigned a rectangle area with their subcategory rectangles nested inside of it.
When a quantity is assigned to a category, its area size is displayed in proportion to that quantity and to the other quantities within the same parent category in a part-to-whole relationship. 
Also, the area size of the parent category is the total of its subcategories. If no quantity is assigned to a subcategory, then its area is divided equally amongst the other subcategories within its parent category.

In this study, we design the tree map to show two visualizations, which have different patterns of categories.
\begin{itemize}
    \item \textbf{File-level visualization}. The file level visualization follows the category pattern as \texttt{Project} $\rightarrow$ \texttt{Filename} $\rightarrow$ \texttt{Competency Score} 
    \item \textbf{Contributor-level visualization}. The contributor level visualization follows the category pattern as \texttt{Project} $\rightarrow$ \texttt{Contributor} $\rightarrow$ \texttt{Competency Score} 
\end{itemize}

Note that the area size refers to the number of elements of competency found in each file.

\begin{figure*}
    \centering
    \subfigure[Request]{\includegraphics[width=0.49\textwidth]{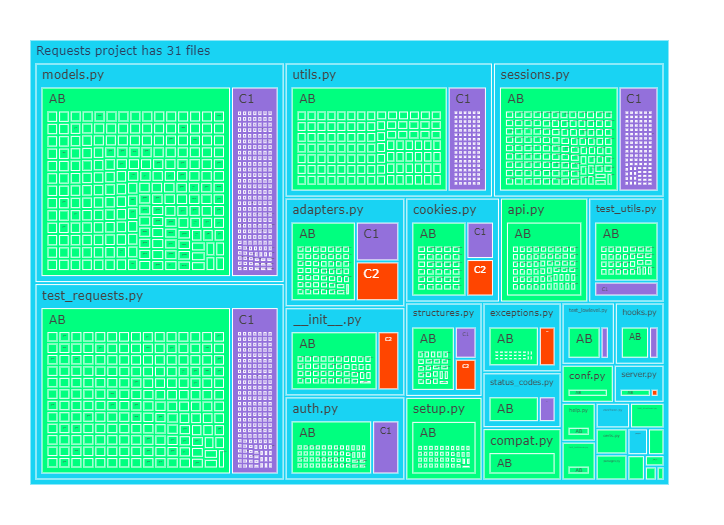}
    \label{fig:requests_file}}
    \subfigure[Flask]{\includegraphics[width=0.49\textwidth]{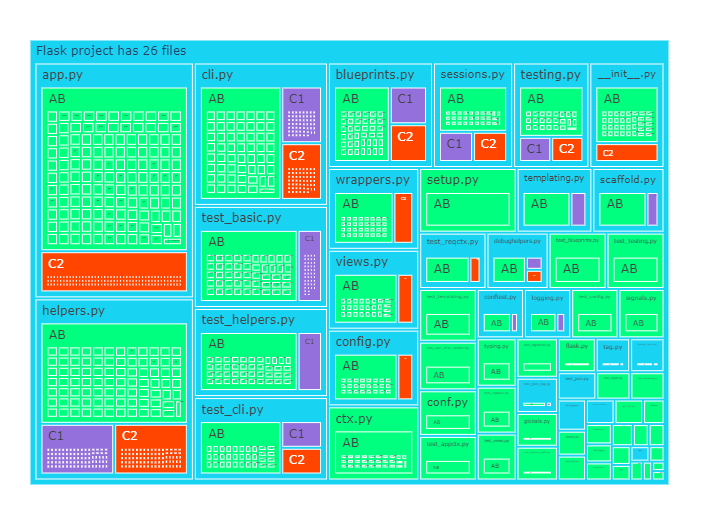}
    \label{fig:flask_file}}
    \subfigure[Jinja2 project]{\includegraphics[width=0.49\textwidth]{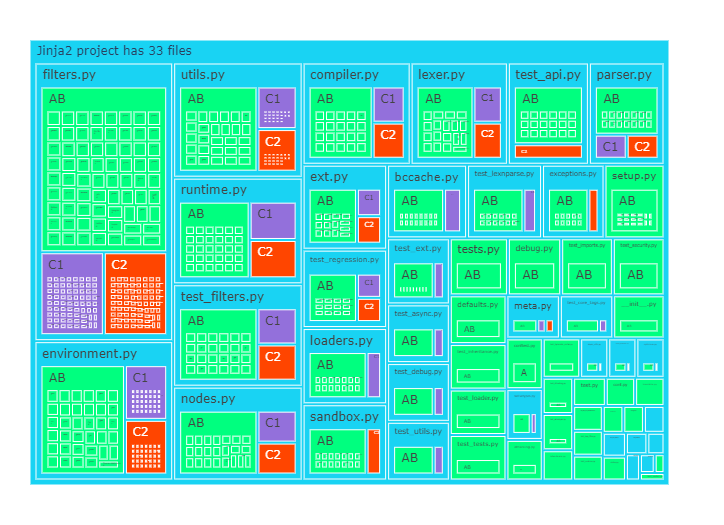}
    \label{fig:jinja_file}}
    \subfigure[Pytz project]{\includegraphics[width=0.49\textwidth]{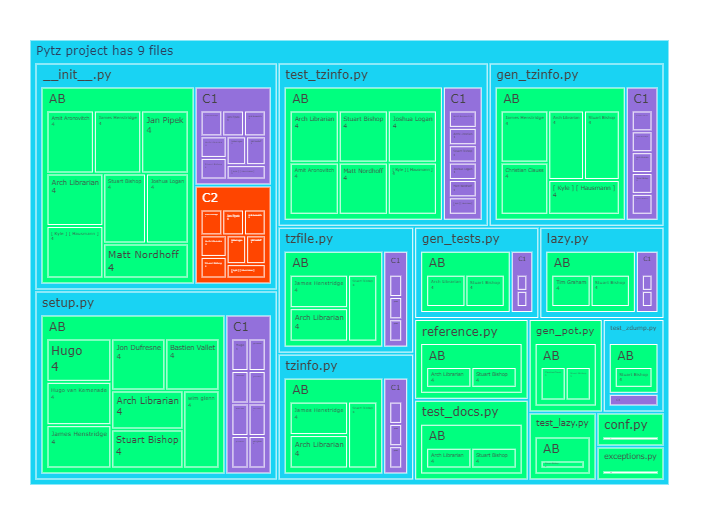}
    \label{fig:pytz_file}}
    \caption{Visualization at the File-Level. \textcolor{green}{green} color represents the {\textit{A and B competency}}, \textcolor{violet}{purple} represents the \textit{C1 - Effective}, and \textcolor{red}{red} represents the C2 - Mastery.}
    \label{fig:Visualizating_at_theFileLevel}
\end{figure*}

\section{A Case Study of Four Projects}
\label{sec:collectdata}
To evaluate our visualization, we performed a case study on four projects. 
\subsection{Data Selection and Collection}


As described in Table~\ref{tab:overviewproject}, we selected our case study based on being the top four Python PyPI libraries (based on the number of dependents to other packages in the PyPI package ecosystem), that are, 1) Requests\footnote{\url{https://github.com/psf/requests}}, an elegant and simple HTTP library for Python, built for human beings.
2) Flask\footnote{\url{https://github.com/pallets/flask}}, a lightweight Web Server Gateway Interface (WSGI) web application framework. It is designed to make getting started quick and easy, with the ability to scale up to complex applications.
3) Jinja2\footnote{\url{https://github.com/pallets/jinja}}, a full-featured template engine for Python. 
4) Pytz\footnote{\url{https://github.com/stub42/pytz}} that allows accurate and cross-platform timezone calculations using Python 2.4 or higher. It also solves the issue of ambiguous times at the end of daylight saving time.

\subsection{Extracting Competency Levels}


For computing the compentency we utilized the \textit{pycefr}~\cite{robles2022pycefr} tool, as it evaluates competency at the file level.
\textit{Pycefr} will analyze all of the files inside the repository and give the results as classified the types of "competency levels", along with "file names" of all files inside each repository - it also contains classified the types of "Repository name, Class, Start Line, End line, and Displacement".
For the four Python PyPI projects,  a total of 149 files.  

\begin{table}
	\centering
	\caption{Level Assigned to an excerpt of Python elements. Taken from \cite{robles2022pycefr}}
	\begin{tabular}{clc}
		\toprule
		No & Python Element & Level \\
		\midrule
		1 & Print & A1 \\
		2 & If statement & A1 \\
		3 & List & A1 \\
		4 & Open function (files) & A2 \\
		5 & Nested list & A2 \\
		6 & List with a dictionary & B1 \\
		7 & Nested dictionary & B1 \\
		8 & with & B1 \\
		9 & List comprehension & B2 \\
		10 & \_\_dict\_\_ attribute & B2 \\
		11 & \_\_slots\_\_  & C1 \\
		12 & Generator function & C1 \\
		13 & Meta-class & C2 \\
		14 & Decorator class & C2 \\
		\bottomrule
	\end{tabular}
	\label{tab:python-elements}
\end{table}

For our classification, we use \textit{pycefr} classification of competency. 
These are \textit{Basic (A)} that refers to code {in which} the beginners could modify the few elements and easy structure of the elementary program,  \textit{Independent (B)} {that refers} to code modified by an experienced programmer that has extended learning in the intermediate program, and \textit{Proficient (C)} that refers to code by an advanced programmer or mastery in acquiring more complex and advanced elements.
Examples of proficiency are shown in Table \ref{tab:python-elements}.

\begin{figure*}
    \centering
    \subfigure[Request]{\includegraphics[width=0.49\textwidth]{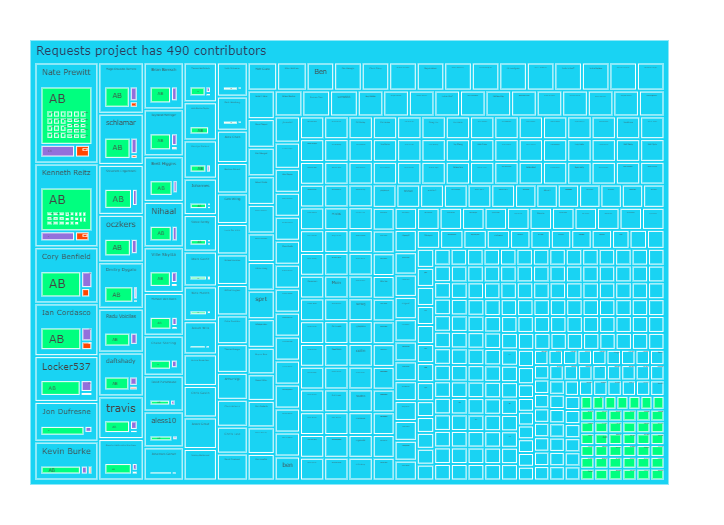}
    \label{fig:requests_cont}}
    \subfigure[Flask]{\includegraphics[width=0.49\textwidth]{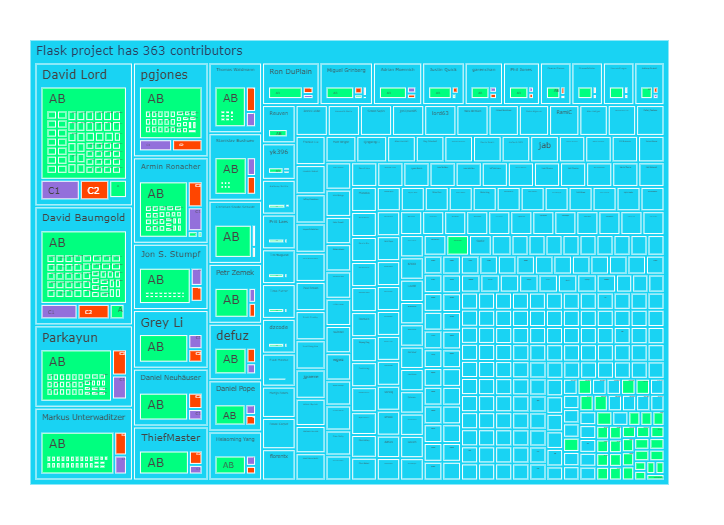}
    \label{fig:flask_cont}}
    \subfigure[Jinja2]{\includegraphics[width=0.49\textwidth]{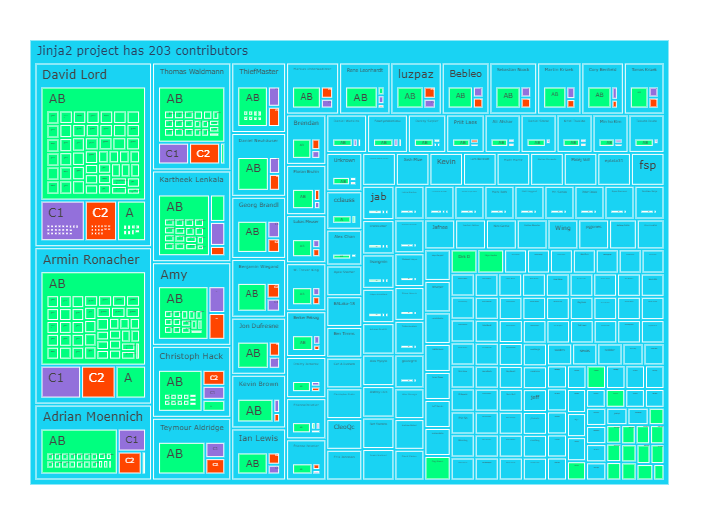}
    \label{fig:jinja_cont}}
    \subfigure[Pytz]{\includegraphics[width=0.49\textwidth]{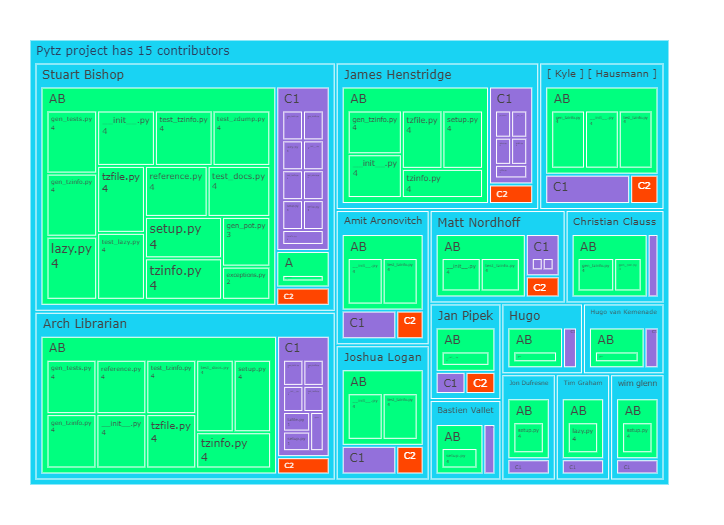}
    \label{fig:pytz_cont}}
    \caption{Visualization at the Contributor-Level. \textcolor{green}{green} color represents the {\textit{A and B competency}}, \textcolor{violet}{purple} represents the \textit{C1 - Effective}, and \textcolor{red}{red} represents the C2 - Mastery.}
    \label{fig:Visualizating_at_theContributorLevel}
\end{figure*}

Since our visualization is used to compare between the basic and proficient, we combine the competency level to form three groups.
\begin{itemize}
    \item \textit{Basic (A \& B)} - This is the code competency that combines the Basic and Independent levels.  
    \item \textit{Proficient (C1) Effective } - This is the effective operational or advanced code competency
    \item \textit{Proficient (C2) Mastery } - This is the mastery or proficiency code competency
\end{itemize}

\begin{table}[]
    \centering
    \caption{ Frequency of file based on the competency.}
    \begin{tabular}{ccccc}
        \toprule
        \multirow{2}{*}{\textbf{Competency Level}} & \multicolumn{4}{c}{\textbf{\# Files for each competency}} \\ 
        \cline{2-5}
        \noalign{\vskip 1mm} 
         & {\textbf{Requests}} & {\textbf{Flask}} & {\textbf{Jinja2}} & {\textbf{Pytz}} \\
        \midrule
        \multicolumn{1}{c}
                    {\textbf{AB (A1, A2, B1, B2)}} & 31 & 62 & 60 & 15 \\
        \multicolumn{1}{c}
                    {\textbf{C1 - Effective}} & 14 & 20 & 28 & 9 \\
        \multicolumn{1}{c}
                    {\textbf{C2 - Mastery}} & 6 & 15 & 19 & 1\\
        \bottomrule
    \end{tabular}
    \label{tab:stats}
\end{table}

\subsection{Mapping Contributor to Competency}

In this analysis, we used the \textit{pydriller}\cite{Spadini2018}.
We started by collecting commits for all python projects and filtering commits made by the authors, along with the filename, which would be useful in identifying where authors do commit. 

Table \ref{tab:stats} shows the mapping the Contributor commits to each file for 5,731 contributors and 842 files.
As shown in the table, we can see that many of the files do contain proficient files (20, 35, 47, and 10 files for the four projects).
Please note that we remove duplication files with the same competency.


\section{Findings}

In this section, we present our visualization from two viewpoints. 
To better understand the visualization, we use different colours to represent the competency.
In this case, the \textcolor{green}{green} color represents the {\textit{A and B competency}}, \textcolor{violet}{purple} represents the \textit{C1 competency}, and \textcolor{red}{red} represents the C2 competency.
Note that the purple and red indicate the more proficient code.
Also, the size of the box indicates that these are many cases of code elements.
We now discuss each visualization.

\paragraph{Competency at the File Level} Figure \ref{fig:Visualizating_at_theFileLevel} shows the competency for each file. In summary, we analyze Jinja project 33 files, Requests 31 files, Flask 26 files, and Pytz  9 files. 
As we can see in all projects, the most prevalent competency is basic competency (A \& B) in green color. 
For example, the Request project has \texttt{models.py} as the files that contain the most code elements.
Furthermore, most of these are AB competencies.
Overall, we can see that for all files, these is abit of elegant code (in purple), but not larger when compared to the basic code elements, which is consist for all four projects.
Thus we make our first observation.

\begin{tcolorbox}
\textbf{Answering RQ1}:
Our visualization shows that Basic Competency (A \& B) is prevalent in PyPI projects, but also contains a much smaller percentage of proficient code.
\end{tcolorbox}

\paragraph{Competency at the Contributor-Level}
Figure \ref{fig:Visualizating_at_theContributorLevel} shows results of the competency of code that is committed by a contributor. 
We analyze Jinja with 203 contributors, Requests with 490 contributors, Flask with 363 contributors, and Pytz with 15 contributors.
Visually, we can see that different contributor make different amounts of contributions. For example, in the Jinja project (Figure 3c), contributor David Lord makes the most contributions, since he has the largest size of the box.
But interestingly, they still commit to files with more elegant code, as shown with the purple color.
Hence, we make our second observation.

\begin{tcolorbox}
\textbf{Answering RQ2}: Our visualization shows that most contributors still contribute basic competency code compared to proficient code. 
\end{tcolorbox}









\section{Limitations}
Since we only focus on four PyPI projects, we acknowledge that generalization is a key threat.
However, as in the initial concept, we are confident our preliminary result can be strengthened with a complete study, which we hope to do in the future.
Another key threat is the granularity of the mapping between competency and the contributor. 
Since one file may be classified as having different competencies,  we cannot determine for sure whether the contributor commits proficient code, instead, they commit to files with that competency.
As an early results paper, our goal was to make preliminary observations, and then improve the accuracy in the future. 
Another threat in our study relates to the readability of our visualization due to the size of the dataset. 
We acknowledge this threat, and plan to carry out a more comprehensive analysis of larger size code-bases of Python projects in the complete study.

\section{Implications and Research Outlook}
\label{sec:conclusion}

We highlight two takeaways and the research outlook.
\paragraph{Every project uses basic code in their files}
Our key takeaway message from RQ1 is that most projects are comprised of very basic and independent code. 
On the other hand, it seems that many files have more proficient code. 
The implication is that the Python community does really do keep the code simple, but some portions of the file being elegant. 
For future work, it would be interesting to look at the actual filename and conduct a deeper analysis to describe why more proficient code is required. 

\paragraph{Not every contributor submits proficient code}
Our key takeaway message from RQ2 is that most contributors do not submit to files that have proficient code. 
Although we need more evidence, the results indicate that proficient code is not needed to make a contribution. 
This is good news for newcomers and those that want to make contributions.
Especially in the context of Pythonic coding, we find that Pythonic or elegant coding is not needed every time. 
Potential future work would be the identification of those experienced contributors, to understand what competency of code they use. 

Potential uses of this visualization are to identify potential experts, and also which files contain more advanced code. 
This could open up new areas of research, especially in terms of maintenance such as bug fixing, new features, and other activities. 

\section*{Acknowledgment}
This work has been supported by 
JSPS KAKENHI Grant Numbers JP8H04094, JP20K19774, and JP20H05706.

\bibliographystyle{IEEEtran}
\bibliography{reference.bib}

\begin{thebibliography}{10}
\providecommand{\url}[1]{#1}
\csname url@samestyle\endcsname
\providecommand{\newblock}{\relax}
\providecommand{\bibinfo}[2]{#2}
\providecommand{\BIBentrySTDinterwordspacing}{\spaceskip=0pt\relax}
\providecommand{\BIBentryALTinterwordstretchfactor}{4}
\providecommand{\BIBentryALTinterwordspacing}{\spaceskip=\fontdimen2\font plus
\BIBentryALTinterwordstretchfactor\fontdimen3\font minus
  \fontdimen4\font\relax}
\providecommand{\BIBforeignlanguage}[2]{{%
\expandafter\ifx\csname l@#1\endcsname\relax
\typeout{** WARNING: IEEEtran.bst: No hyphenation pattern has been}%
\typeout{** loaded for the language `#1'. Using the pattern for}%
\typeout{** the default language instead.}%
\else
\language=\csname l@#1\endcsname
\fi
#2}}
\providecommand{\BIBdecl}{\relax}
\BIBdecl

\bibitem{cock2009biopython}
P.~J. Cock, T.~Antao, J.~T. Chang, B.~A. Chapman, C.~J. Cox, A.~Dalke,
  I.~Friedberg, T.~Hamelryck, F.~Kauff, B.~Wilczynski \emph{et~al.},
  ``Biopython: freely available python tools for computational molecular
  biology and bioinformatics,'' \emph{Bioinformatics}, vol.~25, no.~11, pp.
  1422--1423, 2009.

\bibitem{alexandru2018usage}
C.~V. Alexandru, J.~J. Merchante, S.~Panichella, S.~Proksch, H.~C. Gall, and
  G.~Robles, ``On the usage of pythonic idioms,'' in \emph{Proceedings of the
  2018 ACM SIGPLAN International Symposium on New Ideas, New Paradigms, and
  Reflections on Programming and Software}, 2018, pp. 1--11.

\bibitem{10.1145/3486608.3486909}
\BIBentryALTinterwordspacing
A.~Farooq and V.~Zaytsev, ``There is more than one way to zen your python,'' in
  \emph{Proceedings of the 14th ACM SIGPLAN International Conference on
  Software Language Engineering}, ser. SLE 2021.\hskip 1em plus 0.5em minus
  0.4em\relax New York, NY, USA: Association for Computing Machinery, 2021, p.
  68–82. [Online]. Available: \url{https://doi.org/10.1145/3486608.3486909}
\BIBentrySTDinterwordspacing

\bibitem{robles2022pycefr}
G.~Robles, R.~G. Kula, C.~Ragkhitwetsagul, T.~Sakulniwat, K.~Matsumoto, and
  J.~M. Gonzalez-Barahona, ``pycefr: Python competency level through code
  analysis,'' \emph{arXiv preprint arXiv:2203.15990}, 2022.

\bibitem{phanudom2020teddy}
P.~Phan-Udom, N.~Wattanakul, T.~Sakulniwat, C.~Ragkhitwetsagul, T.~Sunetnanta,
  M.~Choetkiertikul, and R.~G. Kula, ``Teddy: automatic recommendation of
  pythonic idiom usage for pull-based software projects,'' in \emph{2020 IEEE
  International Conference on Software Maintenance and Evolution
  (ICSME)}.\hskip 1em plus 0.5em minus 0.4em\relax IEEE, 2020, pp. 806--809.

\bibitem{Sakulniwat}
T.~Sakulniwat, R.~G. Kula, C.~Ragkhitwetsagul, M.~Choetkiertikul,
  T.~Sunetnanta, D.~Wang, T.~Ishio, and K.~Matsumoto, ``Visualizing the usage
  of pythonic idioms over time: A case study of the with open idiom,'' in
  \emph{2019 10th International Workshop on Empirical Software Engineering in
  Practice (IWESEP)}, 2019, pp. 43--435.

\bibitem{leelaprute2022does}
P.~Leelaprute, B.~Chinthanet, S.~Wattanakriengkrai, R.~G. Kula, P.~Jaisri, and
  T.~Ishio, ``Does coding in pythonic zen peak performance? preliminary
  experiments of nine pythonic idioms at scale,'' \emph{arXiv preprint
  arXiv:2203.14484}, 2022.

\bibitem{zhang2022making}
Z.~Zhang, Z.~Xing, X.~Xia, X.~Xu, and L.~Zhu, ``Making python code idiomatic by
  automatic refactoring non-idiomatic python code with pythonic idioms,''
  \emph{arXiv preprint arXiv:2207.05613}, 2022.

\bibitem{9604836}
D.~P. Tua, R.~Minelli, and M.~Lanza, ``Voronoi evolving treemaps,'' in
  \emph{2021 Working Conference on Software Visualization (VISSOFT)}, 2021, pp.
  1--5.

\bibitem{9373115}
A.~Macquisten, A.~M. Smith, and S.~Johansson~Fernstad, ``Evaluation of
  hierarchical visualization for large and small hierarchies,'' in \emph{2020
  24th International Conference Information Visualisation (IV)}, 2020, pp.
  166--173.

\bibitem{Spadini2018}
\BIBentryALTinterwordspacing
D.~Spadini, M.~Aniche, and A.~Bacchelli, ``{PyDriller: Python framework for
  mining software repositories},'' in \emph{Proceedings of the 2018 26th ACM
  Joint Meeting on European Software Engineering Conference and Symposium on
  the Foundations of Software Engineering - ESEC/FSE 2018}.\hskip 1em plus
  0.5em minus 0.4em\relax New York, New York, USA: ACM Press, 2018, pp.
  908--911. [Online]. Available:
  \url{http://dl.acm.org/citation.cfm?doid=3236024.3264598}
\BIBentrySTDinterwordspacing

\end{thebibliography}


\end{document}